\documentclass{article}
\usepackage[preprint]{spconf}
\usepackage{amsmath,graphicx}
\usepackage{multirow}

\title{Weak-Supervised Dysarthria-invariant Features for
Spoken Language Understanding using an FHVAE and Adversarial Training}
\name{Jinzi Qi$^1$, Hugo Van hamme$^1$}
\address{
  $^1$KULeuven, Department Electrical Engineering-ESAT-PSI,\\Kasteelpark Arenberg 10, Leuven, Belgium}
\usepackage{fancyhdr}
\begin{document}
\thispagestyle{fancy}

\maketitle

\begin{abstract}

The scarcity of training data and the large speaker variation in dysarthric speech lead to poor accuracy and poor speaker generalization of spoken language understanding systems for dysarthric speech.
Through work on the speech features, we focus on improving the model generalization ability with limited dysarthric data. 
Factorized Hierarchical Variational Auto-Encoders (FHVAE) trained unsupervisedly have shown their advantage in disentangling content and speaker representations. Earlier work showed that the dysarthria shows in both feature vectors.
Here, we add adversarial training to bridge the gap between the control and dysarthric speech data domains. We extract dysarthric and speaker invariant features using weak supervision. The extracted features are evaluated on a Spoken Language Understanding task and yield a higher accuracy on unseen speakers with more severe dysarthria compared to features from the basic FHVAE model or plain filterbanks.

\end{abstract}

\begin{keywords}
 Dysarthric speech, FHVAE, adversarial training, weak supervision, end-to-end spoken language understanding.
\end{keywords}

\section{Introduction}

Dysarthric speech data, especially accurately labeled data, are scarce due to difficulties in recruitment, collection and labeling. This data insufficiency problem has been a constant barrier for automatic speech recognition (ASR) model training. Recent ASR models\cite{vaswani2017attention, devlin2018bert, gulati2020conformer} tend to be more complex and employ more parameters, which increases the difficulty to train from limited dysarthric data even further.

Current research mainly solves the dysarthric speech data deficiency problem in three ways. One solution is taking abundant canonical (control) speech data into account during training, e.g. pretraining on control speech and finetuning on dysarthric speech \cite{shor2019personalizing, takashima2019knowledge,wang2021study}. A second approach is to decrease the model size \cite{wang2021light}, or to train an inserted small module instead of finetuning the whole model \cite{tomanek2021residual, biadsy2022scalable}, so the number of parameters learned on the dysarthric data is limited. Thirdly and differently from the solutions that work on training strategy or model structure, \cite{jiao2018simulating, sudro2021significance, jin2021adversarial, jin2022personalized} focus directly on the data and do augmentation to generate more dysarthric speech for use in training. 

The features used in most methods preserve the speaker characteristics and the dysarthria information, and thus bound the model generalization ability. However, in practical scenarios, we would like the ASR model to have great generalization ability and work well on unseen speakers (users), to reduce any training effort for dysarthric users. 
In this work, we focus on improving the automatic recognition model generalization ability with limited dysarthric data. 
Instead of training and testing the model with speech features that preserve variability between dysarthric speakers, we propose to use dysarthric-invariant and speaker independent speech features for automatic recognition use, which can both boost the model generalization ability on unseen speakers and gives the possibility to involve abundant canonical speech into training. 

Factorized Hierarchical Variational Auto-Encoders (FHVAE) \cite{hsu2017unsupervised, shon2018unsupervised} have shown their ability to disentangle the content space and the speaker space in input speech. An FHVAE models the generative process of a sequence of segments in a hierarchical structure. It encodes a speech utterance into a segment-related variable (short time scale) and a sequence-related variable (long time scale) via two linked encoders. The segment-related latent variable represents the information that only appears in a single segment in the sequence, such as the content of the segment, and is conditioned on the sequence-related variable. The sequence-related variable reflects the features of the whole sequence, like the acoustic environment or speaker characteristics. The model hence offers a separation between speaker characteristics (sequence) and content (segment). The content variable (segment-related variable) appears to fit the speaker-independence desideratum outlined above. However, our previous work \cite{qi2021speech} shows that the dysarthria is reflected in both latent variables. 

To achieve a greater deal of dysarthria-invariance in the content variable, instead of using forced regularization with phoneme alignment \cite{qi2021speech}, inspired by \cite{rumberg2021age}, we introduce adversarial training \cite{gui2021review} to the FHVAE model. We feed the FHVAE model with both control and dysarthric speech, regarding the content variable encoder as the generator. We add a discriminator that tries to distinguish control from dysarthric content variables. With adversarial training between the generator and the discriminator, we can force the content variable encoder (generator) to extract content information from both control and dysarthric speech that shares the same dysarthria-invariant latent space. To make the extracted variable closer to the control data space and improve disentanglement, we also try two additional loss terms called reference loss and disentanglement loss (see section 2). Note that we only need the speaker type label as weak supervision of the training of the complete proposed model. The content encoder is used by both dysarthric speech and canonical speech, which gives opportunities to involve abundant canonical speech features into the downstream model (ASR) training in future work. 

For evaluation, we use the trained content encoder to extract the content variables from other datasets and use these variables to train and test a mature End-to-End (E2E) Spoken Language Understanding (SLU) system \cite{renkens2019assist, renkens2018capsule}. The system consists of a recurrent neural network (RNN) encoder, a capsule network \cite{sabour2017dynamic} and output layers, and can transfer input speech features into an intent (semantic) representation without using an explicit textual representation. 
E2E SLU evaluation is different from the evaluation on speaker-independent, in the sense that we have (a limited number of) task-specific recordings by dysarthric speakers at our disposal for training the SLU system, whereas ASR systems would typically be trained on task-agnostic data. Because of the domain adaptation inherent in the evaluation, it is harder to show generalization properties on E2E SLU tasks.

Apart from SLU evaluation, we also train and test a binary speaker type classifier (control/dysarthria) using both content variables and sequence-related variables, and use the classification accuracy to show how the dysarthria information distribution changes after the adversarial training.

We introduce the feature extraction methods, the FHVAE model with adversarial training, reference loss and disentanglement loss, in section 2. Section 3 describes the data and the experimental settings. Results and analysis will be provided in section 4, and section 5 gives conclusions.

\section{Method}
In this section, we introduce the feature extraction methods: the FHAVE model extended with adversarial training, reference loss and disentanglement loss.

\subsection{FHVAE with adversarial training}
Suppose we have $I$ speech data sequences, each sequence $\textbf{X}^i$ containing $N_i$ segments $\textbf{x}^{i,n}$, $n = 1, 2, ..., N_i$. The segment $\textbf{x}^{i,n}$ will be represented by two latent variables $\textbf{z}^{i,n}_1$ and $\textbf{z}^{i,n}_2$ via an FHVAE:
\begin{equation}
    q(\textbf{z}^{i,n}_2|\textbf{x}^{i,n}) 
    = \mathcal{N}(\mathbf{Enc}_{\mu_{\textbf{z}_2}}(\textbf{x}^{i,n}), \mathbf{Enc}_{\sigma^2_{\textbf{z}_2}}(\textbf{x}^{i,n}))
\end{equation}
\begin{equation}
\label{eq2}
    q(\textbf{z}^{i,n}_1|\textbf{x}^{i,n},\textbf{z}^{i,n}_2) 
    = \mathcal{N}(\mathbf{Enc}_{\mu_{\textbf{z}_1}}(\textbf{x}^{i,n},\textbf{z}^{i,n}_2), \mathbf{Enc}_{\sigma^2_{\textbf{z}_1}}(\textbf{x}^{i,n},\textbf{z}^{i,n}_2)
\end{equation}
where $\mathbf{Enc}_{\mu_{\textbf{z}_2}}(\cdot)$ and $\mathbf{Enc}_{\sigma^2_{\textbf{z}_2}}(\cdot)$ are the networks encoding input $\textbf{x}$ to mean and variance of $\textbf{z}_2$,  $\mathbf{Enc}_{\mu_{\textbf{z}_1}}(\cdot)$ and $\mathbf{Enc}_{\sigma^2_{\textbf{z}_1}}(\cdot)$ are the encoders of $\textbf{z}_1$. Following the theory of variational auto-encoders, the latent variables should follow prior distributions:
\begin{equation}
    p(\textbf{z}^{i,n}_1) 
    = \mathcal{N}(\textbf{0}, \sigma^2_{\textbf{z}_1}\textbf{I})
\end{equation}
\begin{equation}
    p(\textbf{z}^{i,n}_2|\boldsymbol{\mu}^i_2) = \mathcal{N}(\boldsymbol{\mu}^i_2, \sigma^2_{\textbf{z}_2}\textbf{I}), \quad
    p(\boldsymbol{\mu}^i_2) 
    = \mathcal{N}(\textbf{0}, \sigma^2_{\boldsymbol{\mu}_2}\textbf{I})
\end{equation}
where $\sigma^2_{\textbf{z}_1},  \sigma^2_{\textbf{z}_2}$ and $ \sigma^2_{\boldsymbol{\mu}_2}$ are chosen upfront and $\textbf{z}^{i,n}_1, \textbf{z}^{i,n}_2$ are i.i.d samples. The prior enforces variables $\textbf{z}^{i,n}_2$ within the same sequence $i$ to be close.

The $\textbf{z}^{i,n}_1$, conditioned on $\textbf{z}^{i,n}_2$ (see equation\ref{eq2}), only contains information with the speech segment $\textbf{x}^{i,n}$, and is regarded as the speaker-independent content variable. The variable $\textbf{z}^{i,n}_2$ reflects the features of the whole sequence $\textbf{X}^i$ and can be seen as speaker variable. The mean of sequence-level variable $\boldsymbol{\mu}^i_2$ can be inferred as:
\begin{equation}
    \tilde{\mu}^i_2 = \frac{\sum^N_{n=1}\mathbf{Enc}_{\mu_{\textbf{z}_2}}(\textbf{x}^{i,n})}{N+\frac{\sigma^2_{\textbf{z}_2}}{\sigma^2_{\boldsymbol{\mu}_2}}}
\end{equation}

Then for the whole sequence $\textbf{X}^i$, the inference model is written as:
\begin{equation}
    q(\textbf{Z}^i_1, \textbf{Z}^i_2, \boldsymbol{\mu}^i_2|\textbf{X}^i) = q(\boldsymbol{\mu}^i_2|\textbf{x}^{i,n})
    \prod^N_{n=1}q(\textbf{z}^{i,n}_1|\textbf{x}^{i,n})q(\textbf{z}^{i,n}_2|\textbf{x}^{i,n})
\end{equation}
where $\textbf{Z}^i_1 = \{ \textbf{z}_1^{i,n} \} ^N_{n=1}$, $\textbf{Z}^i_2 = \{ \textbf{z}_2^{i,n} \} ^N_{n=1}$. 
The generative model for the data is then:
\begin{equation}
    p(\textbf{x}^{i,n}|\textbf{z}^{i,n}_1, \textbf{z}^{i,n}_2) 
    = \mathcal{N}(\mathbf{Dec}_{\mu_x}(\textbf{z}^{i,n}_1, \textbf{z}^{i,n}_2), \mathbf{Dec}_{\sigma^2_x}(\textbf{z}^{i,n}_1, \textbf{z}^{i,n}_2))
\end{equation}
where $\mathbf{Dec}_{\mu_x}(\cdot)$ and $\mathbf{Dec}_{\sigma^2_x}(\cdot)$ form the decoder networks of the FHVAE.

The dysarthria information, as a speaker characteristic, should be extracted to speaker variable $\textbf{z}^{i,n}_2$ by the FHVAE. However, our previous work \cite{qi2021speech} found that the FHVAE does not separate the dysarthria and content information and speech impairment is identifiable from $\textbf{z}^{i,n}_1$.

\begin{figure}[htbp]
  \centering
  \includegraphics[width=3.2in]{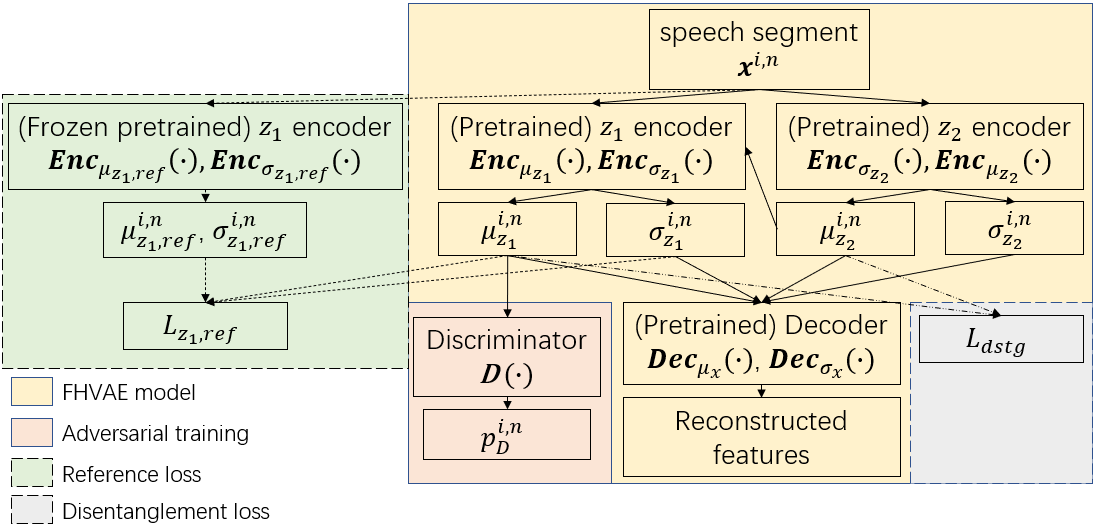}
  \caption{Dataflow in the proposed model with adversarial training, reference loss and disentanglement loss.}
  \label{fig:dataflowchart}
\end{figure}

To obtain dysarthria-invariant features $\textbf{z}^{i,n}_2$, inspired by \cite{rumberg2021age}, we introduce adversarial training into the FHVAE model. The data flow of the proposed model is provided in figure \ref{fig:dataflowchart}.

Repurposing the $\mathbf{z}_1$ mean encoder $\mathbf{Enc}_{\mu_{\textbf{z}_1}}(\cdot)$ as the generator in adversarial training, we get $\mu_{\textbf{z}_{1}}^{i,n}$ as the generator output and feed this into a discriminator $\mathbf{D}(\cdot)$ which is trained to yield 
the probability $p^{i,n}_{D}$ that a speech segment $\textbf{x}^{i,n}$ stems from a control speaker:
\begin{equation}
    p^{i,n}_{D} = \mathbf{D}(\mu_{\textbf{z}_{1}}^{i,n}) = \mathbf{D}(\mathbf{Enc}_{\mu_{\textbf{z}_1}}(\textbf{x}^{i,n}))
\end{equation}
Training of $\mathbf{D}(\cdot)$ involves dysarthric and control data and minimizes cross-entropy:
\begin{equation}
    L_{D} 
    = CE_{dys} + CE_{ctrl}
\label{eq:dloss}
\end{equation}
where
\begin{equation}
\begin{split}
CE_{dys/ctrl} =\mathrm{E}_{\mu_{\textbf{z}_1}\sim p_{dys/ctrl}({\mu_{\textbf{z}_1})}}[
-(l^{i,n}_{dys/ctrl} \times\log p^{i,n}_{D})&\\
+(1-l^{i,n}_{dys/ctrl}) \times\log (1-p^{i,n}_{D})]&
\end{split}
\end{equation}
and $l^{i,n}_{dys/ctrl}$ is the label of speech segment $\textbf{x}^{i,n}$, i.e. $l^{i,n}_{dys/ctrl}$ is $1$ when $\textbf{x}^{i,n}$ is dysarthric and is $0$ for control speech. 

The loss function of the FHVAE model can be written as:
\begin{equation}
\label{eq:ltot}
    L_{FHVAE} = L_{LB} + L_{z_2,disc} * W_{z_2,disc} + L_{Gen} * W_{Gen}
\end{equation}
where $L_{LB}$ is the variational lower bound loss of the FHVAE model that guarantees the performance of the FHVAE model as an auto-encoder; $L_{z_2, disc}$ is the discriminative loss of the sequence-related latent variable, which expresses that sequence identity should be inferrable from the sequence-related embedding vector, i.e., it is used to encourage separation in the latent space. $L_{Gen}$ is the cross-entropy between the discriminator output and the opposite dysarthric/control label (control is $0$ and dysarthric is $1$), similar to equation \ref{eq:dloss}. Unlike a classical Generative Adversarial Network (GAN), the $L_{Gen}$ here acts on the whole FHVAE model instead of only on the generator part. 

\subsection{Extension with reference loss and disentanglement loss}

In the model presented so far, the adversarial training will modify the encoder output on both control and dysarthric input, as we want a uniform feature extractor for both types of speech. Modifying the behavior for control data is not desirable since the FHVAE is supposed to be pre-trained well in that part of the input space. Thus we add an additional loss term $L_{z_1,ref}$ called reference loss into the FHVAE model training loss with weight $W_{z_1,ref}$. The reference loss is active on control speech only and is the Kullback-Leibler divergence between the reference $z_1$ distribution $\mathcal{N}(\mu_{z_1,ref}, \sigma^2_{z_1,ref})$ of the FHVAE pre-trained on control speech and the $z_1$ distribution during training. We also consider evaluating the $L_{Gen}$ term in Eq. \ref{eq:ltot} only on the dysarthric data ($CE_{dys}$), while the other terms are unchanged. In doing so, the content encoder emphasizes changes for dysarthric speech as well. 

Finally, to further disentangle speaker and content information, we add the disentanglement loss $L_{dstg}$ with weight $W_{dstg}$, which minimizes the sum of squares of the entries of the correlation matrix between $\mu_{z_1}$ and $\mu_{z_2}$.

The reference loss module and disentanglement loss module are shown in figure \ref{fig:dataflowchart}.

\section{Experiments}

This section explains how we implement and evaluate the proposed model, the datasets and the experimental details.

\subsection{Datasets}
We use four datasets for pretraining, finetuning and evaluation, respectively:
\begin{itemize}
\item \textit{\textbf{CGN} dataset} \cite{oostdijk2002experiences}: we use the 'k' and 'o' subsets from the Dutch CGN dataset for FHVAE model pretraining. It contains 35.2-hours of broadcast news and read speech from 326 control Dutch speakers. 

\item \textit{\textbf{Torgo} dataset} \cite{rudzicz2012torgo} is used for model finetuning, containing 6.3 hours of English speech from 7 control speakers (3.8~h) and 8 speakers with dysarthria (2.5~h).

\item \textit{\textbf{Domotica} dataset} \cite{ons2014self} is used for SLU evaluation, containing 8.8~h Dutch speech from 15 speakers with dysarthria (2 speakers have no intelligibility score and are excluded), uttering commands related to home automation, such as "turn on the light in the kitchen". The commands are encoded to 22 slot value labels and 4 categories, and each category command is represented as 1 or 2 active labels. There are in total 3789 speech commands. The severity level (intelligibility score) of dysarthric speakers can be found in \cite{ons2014self}.

\item \textit{\textbf{COPAS} dataset} \cite{copas} is used for speaker type classification evaluation, containing 7 hours of Dutch speech from 131 control speakers (3~h) and 230 speakers with speech disorders (4~h). 4699 utterances in total with an average duration of 6~seconds.
\end{itemize}

\subsection{Experimental settings}

\begin{figure}[htbp]
  \centering
  \includegraphics[width=3in]{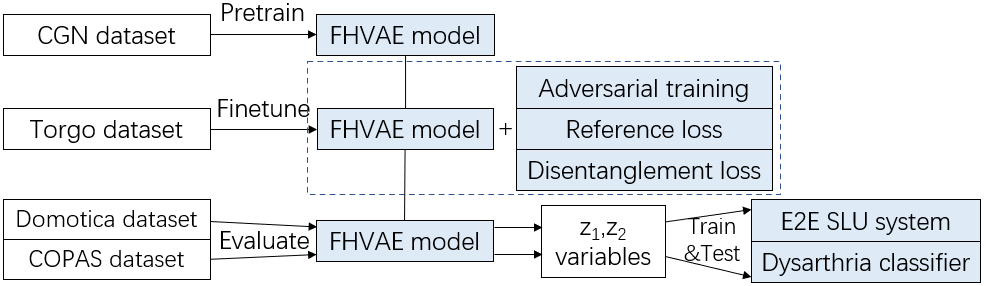}
  \caption{Experimental design.}
  \label{fig:expflowchart}
\end{figure}

We first pre-train the FHVAE model, then finetune it using the adversarial training module and the two additional loss terms. We also provide a comparison where we finetune the pretrained FHVAE without adversarial training.  
Then we use the trained encoders $\mathbf{Enc}_{\mu_{\textbf{z}_1}}(\cdot)$ and $\mathbf{Enc}_{\mu_{\textbf{z}_2}}(\cdot)$ to extract content and sequence-related variables for each segment. The content-related variables are then used to train and test the SLU system, while the speaker type classifier uses both. The experimental design is illustrated in figure \ref{fig:expflowchart}.

\subsubsection{Evaluation setup and metrics}

Since our SLU data contains only speakers with dysarthria, we perform generalization tests across different severity levels rather than control versus dysarthric speech.

The evaluation on SLU uses two train/test divisions. For the Out-of Domain evaluation, we use speech data ($2642$ speech utterances) from 11 speakers with mild to medium dysarthria (intelligibility score $\ge 70$) to train the SLU model and use 4 speakers with severe dysarthria (intelligibility score $< 70$) for testing. 
In this experiment, we evaluate the generalization ability of the model across different severity levels. If the content variable is more dysarthria-invariant, the SLU model should perform better thanks to better generalization. 

In the In-Domain type evaluation, we order the speakers by their intelligibility score from high to low, and use the data ($1901$ speech utterances) from the 8 odd-numbered speakers for training and the rest for testing. In this type, the SLU model sees a wide range of severity levels but tests on unseen speakers within the data domain. 

For SLU system evaluation, the micro-averaged F1-score of the detected labels is used as the metric. For each type, we repeat the experiment 5 times and use the average F1 score as the final result.

For the dysarthria classification, we do 6-fold cross-validation experiment that we divide all COPAS speakers into 6 blocks, each with the same number of control and disordered speech utterances. For each fold, we use four blocks to train the classifier, one for validation and one for testing. We repeat the 6-fold experiment 5 times and use the averaged classification accuracy (correct rate) as the final classification results.

\subsubsection{Network and training setup}

The speech features used in the FHVAE model are $80$ MEL filter bank energies with a frame advance of 10~ms. During training, each FHVAE segment is 20 frames with 8 frames shift, i.e., 200~ms of speech. At inference, the shift is one frame.

In the FHVAE network, the encoder consists of 2 layers of bidirectional LSTM layers and 2 parallel dense layers for each latent variable. Each LSTM has 256 units, and the dimension of each latent variable is 32. The decoder consists of 2 bidirectional LSTM layers of the same size and 2 dense layers of 80 units for $\mu_x$ and $\sigma^2_x$ prediction, respectively. The discriminator consists of two dense layers with 32 hidden nodes and 1 hidden node, respectively. The first layer has $leaky\,ReLU$ activation and a $sigmoid$ is used for the discriminator output. 

We choose the Adam optimizer for training with loss weights $W_{z_2,disc}=10$, $W_{Gen}=500$, $W_{z_1,ref}=0.1$, $W_{dstg}=1$ to match the magnitude of other losses in the final loss function. The hierarchical sampling size is $5000$ \cite{hsu2018scalable}, the batch size is $500$ and the maximal number of training epochs is $50$. Early stopping with patience of $10$ is used during baseline training and simple finetuning. For adversarial training, training stops when it reaches the maximal number of epochs. The default learning rate for the FHVAE model (pre)training and finetuning is $0.001$. For the proposed model with adversarial training, we use a learning rate of $0.001$ for the FHVAE model and $0.0002$ for the discriminator.

We choose the SLU system of \cite{renkens2018capsule} with unmodified setting for evaluation. We set the number of epochs to $50$, use Adam Optimizer with a learning rate $0.001$ and use batch size $16$. For dysarthria classification, the classifier consists of 2 dense layers with 100 and 2 hidden nodes, respectively. The first layer has $ReLU$ activation and the second has $Softmax$ activation. We use a batch size of $128$. We aggregate the input data over the utterance by taking the mean vector as the classifier input data.

\begin{table*}[th]
\caption{F1-scores of SLU tasks (left of the table) and accuracy of dysarthria classifier (right part of the table) using extracted content variables, speaker variables or $Fbank$ features as training and testing data. ($^*: p-value\le0.05$ compared to results of data extracted by FHVAE simply finetuned on Torgo, $^\dagger: p-value\le0.05$ compared to results of $Fbank$ features. The $p-value$s are calculated by Wilcoxon signed rank test on recognition accuracy, where the score per utterance is averaged over 5 trials.)}
\centering
 \label{tab:sum}
\begin{tabular}{llcccc|cccc}
\hline
\multicolumn{2}{l}{\multirow{3}{*}{\textbf{Feature Extraction method}}} & \multicolumn{4}{c|}{\textbf{SLU Task Input}} & \multicolumn{4}{c}{\textbf{Speaker Type Classifier Input}} \\ 
\cline{3-10} 
&& \multicolumn{2}{c}{Out-of-Domain} & \multicolumn{2}{c|}{In-Domain}  &\multirow{2}{*}{$Fbank$} & \multirow{2}{*}{$z_1$} & \multirow{2}{*}{$z_2$} &\multirow{2}{*}{$z_{12}$} \\ 
\cline{3-6}
&& $Fbank$ & $z_1$ & $Fbank$ & $z_1$ &&&& \\ 
\hline
-&&0.526&-&0.576*&-&0.719&-&-&-\\
\multicolumn{2}{l}{FHVAE pretrained on CGN}         
        &-&0.519&-&0.651$^\dagger$&-&0.734$^\dagger$&0.693*&0.757$^\dagger$\\
        & \quad+Simply Finetuned on Torgo   
        &-&0.522 &-&0.653$^\dagger$&-&0.742$^\dagger$&0.680$^\dagger$&0.756$^\dagger$\\
        & \quad+Adversarial training on Torgo
        &-&0.520&-&0.638$^\dagger$&-&0.724*&0.701*&0.755$^\dagger$\\
        & \qquad+Reference loss
        &-&0.537*&-&0.641$^\dagger$&-&0.714*&0.726*&0.754$^\dagger$\\
        & \quad\qquad+Only use $CE_{dys}$ in $L_{Gen}$
        &-&0.571*$^\dagger$&-&0.651$^\dagger$&-&0.720*&0.726*&0.757$^\dagger$\\
        & \qquad\qquad+Disentanglement loss
        &-&0.493*$^\dagger$&-&0.600*$^\dagger$&-&0.707*&0.737*$^\dagger$&0.764$^\dagger$\\
\hline
\end{tabular}                       
\end{table*}

\section{Results and Discussion}
\subsection{E2E SLU system evaluation}

We evaluate how dysarthria and speaker invariant the extracted content variable is.

We first compare SLU performance using filter bank ($Fbank$) versus $\mu_{z_1}$-features extracted by the baselines (pretrained model and simply finetuned model) and proposed models. In table \ref{tab:sum}, we provide the averaged slot value F1-score for the SLU task using different training sets. 

For the Out-of-Domain experiment, the $z_1$ extracted by the FHVAE model with adversarial training has no advantage over the $Fbank$ features when testing the generalization ability between different dysarthria severity levels. The reason might be that the adversarial training strategy successfully narrows the gap between the dysarthric and canonical speech and it does so by finding common ground in which some of the detail is lost for speakers with high intelligibility, leading to accuracy loss in SLU. 
 
By adding the reference loss and including only $CE_{dys}$ in $L_{Gen}$, the modification of the control content space is weakened and the dysarthric speech content space moves closer to the control space. Due to the dysarthria invariance of the content variable, the F1-score increases and we obtain the highest F1-score $0.571$ from the model with adversarial training, reference loss and $CE_{dys}$ only. 
In In-Domain SLU, the baseline FHVAE already outperforms the $Fbank$ features by a great margin. The content variable seems to be a powerful speaker-independent speech representation, a point also made in \cite{hsu2017unsupervised}. For the In-Domain task, it is however less important to narrow the gap between canonical and dysarthric data, since the SLU backend sees both kinds of data in training. The proposed methods for narrowing the gap between canonical and dysarthric speech are hence less relevant in this scenario. However, for the final design with adversarial training, reference loss and restricting the $L_{Gen}$ term to the dysarthric data, there is also no degradation. Hence, the design provides reliable features for SLU in both scenarios. 

Finally, the disentanglement loss seems to hurt the content representation and leads to lower accuracy for SLU.

\subsection{Dysarthria classification}

To further assess the effect of adversarial training in generating a dysarthria-invariant latent variable, we implement a binary classifier that classifies input utterance-level mean vectors into either dysarthric or control type. In the right of table \ref{tab:sum}, we provide the classification accuracy using the content variable $z_1$, speaker variable $z_2$ and $z_{12}$ which concatenates both variables. 

From the table, we see that when finetuning on the dysarthric dataset Torgo, although the F1-score increases, the classification accuracy of $z_1$ increases with an accuracy decrease using $z_2$. The dysarthria information reassignment shows that when processing dysarthric data, the FHVAE tends to extract dysarthria-related information into the content variable $z_1$. When we include the adversarial training into the finetuning procedure, we see a drop in accuracy when using $z_1$ while an increase with $z_2$. This shows that the adversarial training strategy does increase the invariance of the dysarthric information in $z_1$, narrows the gap between dysarthric and control latent spaces and promotes disentanglement of the speaker and content space. Meanwhile, when using both content and speaker variable, classification accuracy with $z_{12}$ is higher than using only $Fbank$ features, showing the potential of the FHVAE model in dysarthric speech diagnosis. 

When we engage the disentanglement loss into model training, there is even lower classification accuracy with $z_1$ and higher accuracy with $z_2$. Hence, the technique is successful in the goal to push dysarthria-related information into the sequence variable. However, this seems to happen at the expense of the quality of the content variable (see section 4.1). Indeed, a lower correlation can also be achieved by making the content variable more noisy.

\section{Conclusions}

We propose a dysarthria-invariant and speaker-independent feature extraction method using an FHVAE model and adversarial training. We work in a setting of limited task-agnostic dysarthric speech data for building the feature extractor and evaluate - again on limited task-specific training data - on an SLU task. By repurposing the content encoder in the FHVAE as a generator and adding a discriminator network, we are able to narrow the gap in the latent space for dysarthric and control speech, using only weak supervision. Using the extracted content variable to train and test a mature E2E SLU system, we see significant F1-score improvement on speakers with more severe dysarthria, compared to using filter bank features or representations from a basic FHVAE model. The binary dysarthria/control classifier also yields lower accuracy when using the extracted content variable, showing the effect of the proposed model in disentangling speaker and content information in dysarthric speech. Interestingly, the extra disentanglement loss does improve disentanglement, but it seems to spoil the content information as well. 

We test the effectiveness of the extracted dysarthria-invariant and speaker-independent features across different dysarthria severity levels. In future work, we intend to evaluate the proposed features of both dysarthric and control speech in other speech recognition models. Moreover, we see obvious improvement in In-Domain evaluation when using pretraining and finetuning from our current work.

\section{Acknowledgements}
The research was supported by KU Leuven Special Research Fund grant C24M/22/025 and the Flemish Government under the ”Onderzoeksprogramma Artificiële Intelligentie (AI) Vlaanderen” programme.

\bibliographystyle{IEEEbib}
\bibliography{mybib}

\end{document}